\documentclass{vgtc}                          

\let\ifpdf\relax

\usepackage{mathptmx}
\usepackage{graphicx,graphics}
\usepackage{times}
\usepackage[usenames,dvipsnames]{color}
\usepackage{subfigure}
\usepackage{amsmath, amsthm, amssymb}
\usepackage[lined, figure]{algorithm2e}
\usepackage{enumerate}
\usepackage[utf8]{inputenc}
\usepackage[hyphens]{url}
\usepackage{float}
\usepackage{flushend}
\usepackage[table,usenames,dvipsnames]{xcolor}
\usepackage{booktabs}
\usepackage{multirow}
\usepackage[10pt]{moresize}
\usepackage[table]{xcolor}

\usepackage[bookmarks,backref=true,linkcolor=black]{hyperref} 
\hypersetup{
  pdfauthor = {},
  pdftitle = {},
  pdfsubject = {},
  pdfkeywords = {},
  pdfstartview={FitH},
  colorlinks=true,
  linkcolor= black,
  citecolor= black,
  pageanchor=true,
  urlcolor = black,
  plainpages = false,
  linktocpage
}

\newcommand{\etal}{{\it et~al.}\ }
\newcommand{\eg}{e.g.}

\allowdisplaybreaks
\onlineid{0}

\vgtccategory{Research}

\vgtcinsertpkg

\title{Augmenting Mobile Phone Interaction with Face-Engaged Gestures}

\author{
Jian Zhao\thanks{e-mail: jianzhao@dgp.toronto.edu} \and%
Ricardo Jota\thanks{e-mail: jotacosta@dgp.toronto.edu} \and%
Daniel Wigdor\thanks{e-mail: dwigdor@dgp.toronto.edu} \and%
Ravin Balakrishnan\thanks{email: ravin@dgp.toronto.edu}
}
\affiliation{\scriptsize \centering Department of Comptuer Science, University of Toronto}

\abstract{
The movement of a user’s face, easily detected by a smartphone’s front camera, is an underexploited input modality for mobile interactions. We introduce three sets of face-engaged interaction techniques for augmenting the traditional mobile inputs, which leverages the combination of the head movements with touch gestures and device motions, all sensed via the phone’s built-in sensors. We systematically present the space of design considerations for mobile interactions using one or more of the three input modalities (i.e., touch, motion, and head). The additional affordances of the proposed techniques expand the mobile interaction vocabulary, and can facilitate unique usage scenarios such as one-hand or touch-free interaction. An initial evaluation was conducted and users had positive reactions to the new techniques, indicating the promise of an intuitive and convenient user experience.
} 

\CCScatlist{
  \CCScat{H.5.2}{[Information Interfaces and Presentation}{User Interfaces}{Interaction Styles}
}

\begin{document}


\firstsection{Introduction}

\maketitle

In recent years, mobile devices, such as smartphones, have become increasingly popular in our everyday life. Touch input, including tapping and flicking, is currently the leading interaction mechanism for high-end mobile phones. However, there are many situations where touch is limited. For instance, when outside during a cold winter, due to limitations of capacitive sensors, users have to take off their gloves to touch the screen, e.g., in order to change the playback of songs; when one hand is otherwise encumbered, users have trouble performing zoom (pinch) actions on their phones, e.g., in navigating maps. Under these circumstances, users could benefit from mechanisms that augment the touch input, although they may not use the augmented gestures all the time.

Modern smartphones support many touch gestures, but also incorporate a myriad of sensors, including accelerometers, gyroscopes, and cameras, that can enable additional interaction affordances. There have been some previous attempts at leveraging these sensors to augment traditional touch input with device motion gestures (e.g., shaking and swiping)~\cite{Bartlett2000,Hinckley2011,Li2009,Roudaut2009}. However, there is another input channel---the movements of a user’s face, which are easily detected by a phone’s front-camera, because people spend much time looking at their phones while using them, yet they are currently not effectively utilized as an input source.

With fast hardware enabling real-time face tracking on mobile devices, several techniques have been proposed to use such information to support more natural interaction, such as auto-screen rotation~\cite{Cheng2012} and multi-perspective panoramas~\cite{Joshi2012}. However, in the previous work, the head is typically used as a static reference to compute the relative screen orientation of the phone, rather than as a highly interactive input modality. Although some off-the-shelf techniques, such as smart scroll~\cite{sumsung}, have also explored the use of head gestures, they are limited in utilizing just the head input channel alone. In contrast, we believe that there is significant potential in leveraging the rich space of a user’s head movements to enhance the expressiveness of existing touch and/or motion gestures when designing novel interaction techniques.

\begin{figure}[!b]
\centering

\includegraphics[width=0.9\columnwidth]{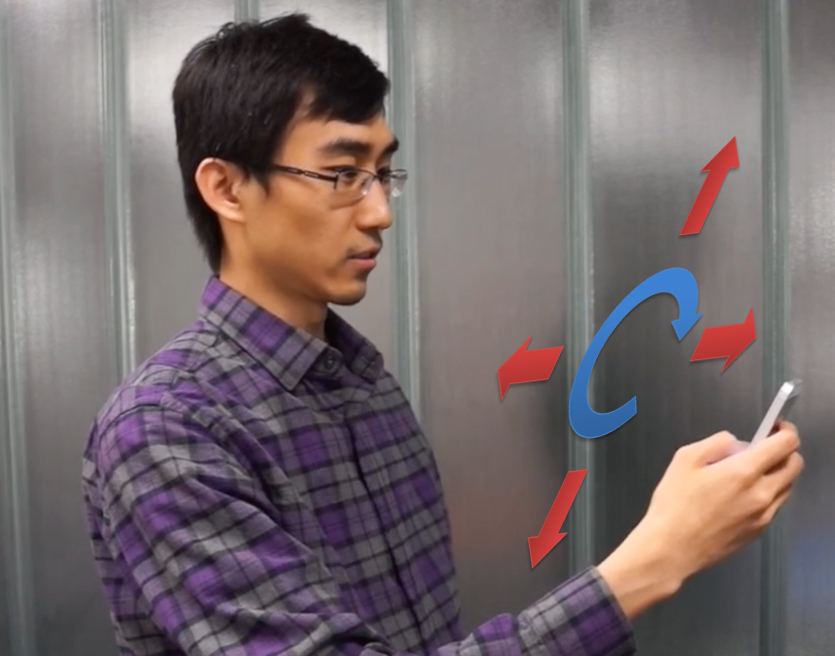}

\caption{Possible head movements when front facing.} \label{fig:example}

\end{figure}

As a first step, we explore such techniques for a subset of all possible movements of a user’s head, i.e., translations and rotations of screen content so that it remains right-side up and in front of the eyes, while the head or phone are moved (Figure~\ref{fig:example}). We focus on face-engaged input methods that can be detected by the built-in sensors of a phone, including the touch sensor, accelerometer, gyroscope, and camera. It is possible that other types of head motions (e.g., turning of the face) or combinations of body parts movements (e.g., whole body and midair hand gestures) can be involved. However, they require either extra hardware (e.g., infrared markers) or are computationally heavy in order to correctly distinguish computer vision features (rather than human front-faces, for which there are highly optimized detectors, in some cases built-in to the handset).

In this paper, we propose three sets of novel face-engaged interaction techniques for mobile phones by combining a user’s head movement with existing input methods (i.e., touches and device motions). We also present a space of design considerations for mobile interactions using the three input channels, which frames the existing methods and our techniques  in a systemic manner, shedding a light for future research. A basic system implementation is described, for tracking a user’s face and fusing multiple sensors to detect gestural inputs with combinations of the three input modalities. The face-engaged techniques can provide extra affordances and improve many of our everyday uses of mobile devices, offering a secondary input approach when necessary. We demonstrate these techniques with a number of common mobile phone interaction tasks, such as scrolling, document browsing, map navigation, and menu selection. As a start of exploring the rich space to leverage face gestures for augmenting traditional inputs, we believe that face-engaged interactions offer a convenient and natural user experience, which could be very useful under some special situations (e.g., no-touch and one-hand).

\section{Design Considerations}
Our goal is to add face movements to traditional touch or motion gestures to augment the current mobile input vocabulary. We thus consider the interaction design systemically based on how the three input modalities, i.e., touch, motion, and face, are used for controlling parameters in mobile interaction, including \emph{discrete control}, \emph{continuous control}, and \emph{not used}~\cite{Hinckley2012}. We define discrete control as mapping the sensor input into discrete events, or modes, such as menu triggering~\cite{Li2009}. Continuous control indicates real-time updates of the interface, based on continuous sensor input, such as scrolling with tilting angles~\cite{Harrison1998}.

As shown in Table~\ref{tab:design}, we frame previous work and techniques proposed in this paper with the above design considerations of the parameter control methods. The first column of Table~\ref{tab:design} groups techniques sharing the same attributes; some techniques may contain one or more attributes, given that there are multiple modes of controlling parameters both continuously and discretely. The proposed face-engaged techniques are highlighted in bold font, which are deliberately chosen to illustrate open opportunities and populate this space, and grouped into three sub-categories, face-engaged touch, face-engaged motion, and face-engaged touch \& motion. As a start, we focus on a subset of all possible head movements (Figure~\ref{fig:example}), and the example techniques only demonstrate certain aspects of using the face gesture to augment traditional mobile input methods. However, Table~\ref{tab:design} presents a clear classification of the techniques, and suggests a much wider space for future mobile interaction design where other degrees of face input can be substantially explored.

\begin{table}[!tb]
\centering
\small
\setlength{\tabcolsep}{3pt}
\begin{tabular}{p{5.5cm}|cc|cc|cc}
\toprule
\multirow{2}{*}{\bf Technique(s)} & \multicolumn{2}{c|}{\bf Face} & \multicolumn{2}{c|}{\bf Motion} & \multicolumn{2}{c}{\bf Touch}  \\ \cmidrule{2-7}
& D & C & D & C & D & C \\ \midrule[\heavyrulewidth]
Rubbing~\cite{Olwal2008}, Microrolls~\cite{Roudaut2009a} & & & & & X & \\ \midrule
Offset Cursor~\cite{Potter1988}, Shift~\cite{Sears1991} & & & & & & X \\ \midrule
TapTap and MagStick~\cite{Roudaut2008} & & & & & X & X \\ \midrule[\heavyrulewidth]
Virtual Shelves~\cite{Li2009}, Auto screen rotation~\cite{Hinckley2000}, Whacking gestures~\cite{Hudson2010}, PhoneTouch~\cite{Schmidt2010} & & & X & & & \\ \midrule
Rock`n’scroll~\cite{Bartlett2000} & & & & X & & \\ \midrule
TimeTilt~\cite{Roudaut2009} & & & X & X & & \\  \midrule[\heavyrulewidth]
TiltText~\cite{Wigdor2003}, Chucking~\cite{Hassan2009}, Wrist angles~\cite{Rahman2009} & & & X & & X & \\ \midrule
Tilt Scrolling~\cite{Harrison1998}, Boom Chameleon~\cite{Tsang2003}, Expressive typing~\cite{Iwasaki2009} & & & & X & X & \\ \midrule
ScatterDice Mobile~\cite{Thomason2012}, TouchProjector~\cite{Boring2010}, Spilling~\cite{Olsen2007} & & & & X & X & X \\ \midrule
Sensor Synaesthesia~\cite{Hinckley2011} & & & X & X & X & X \\ \midrule[\heavyrulewidth]
iOS 7 switch control~\cite{ios7}, Smart stay and Smart pause~\cite{sumsung} & X & & & & & \\ \midrule
Gaze scrolling~\cite{Kumar2007}, Image zoom viewer~\cite{Hansen2006}, Smart scroll~\cite{sumsung} & & X & & & & \\ \midrule[\heavyrulewidth]
{\bf Multi-scale scrolling} & & X & & & & X \\ \midrule
{\bf Coarse-to-fine text edit} & X & & & & X & X \\ \midrule[\heavyrulewidth]
iRotate~\cite{Cheng2012}, Smart rotate~\cite{sumsung} & X & & X & & & \\ \midrule
{\bf 3D map viewer} & & X & X & & & \\ \midrule
Panorama viewing~\cite{Joshi2012}, Smart scroll (tilting device)~\cite{sumsung} & X & & & X & & \\ \midrule
{\bf Touch-free menu} & & X & & X & & \\ \midrule[\heavyrulewidth]
{\bf Expressive flicking} & X & & X & & X & X \\ \midrule
{\bf One-hand navigator} & X & X & & X & X & X \\
\bottomrule
\end{tabular}
\vspace{-2mm}
\caption{Design considerations of mobile interaction based on how touch, motion, and face are used for parameter controls (D: discrete control, C: continuous control, and shaded empty: not used). } \label{tab:design}

\end{table}

\section{Related Work}
In this section, we discuss a summary of relevant mobile interaction techniques, followed by above the design considerations in Table~\ref{tab:design}.

For touch only input techniques, one of the major issues is the imprecision and occlusion of touch interfaces, also known as the “fat finger” problem. For example, Offset Cursor displays the cursor at a fixed distance on the screen to avoid the finger occlusion~\cite{Potter1988}; similarly, Shift applies a hybrid approach where a normal coarse direct touch selection is followed by a precise position adjustment~\cite{Sears1991}. Along this line, Roudaut \etal proposed TapTap and MagStick that further improve the efficiency and accuracy of small target selection~\cite{Roudaut2008}. Another relevant area is that concerning one-handed interaction. Karlson \etal have shown many cases in which users prefer one-handed use for mobile devices~\cite{Karlson2008}. Some gestures for performing one-handed zooming have been widely used on commercial systems, such as “double tap” on iOS and “tap then slide” in Google Map mobile app. Other techniques such as “rubbing”~\cite{Olwal2008} or “rolling”~\cite{Roudaut2009a} have also been proposed to facilitate single hand touch manipulations.

However, touches are not always convenient for a user. Examples include when one is wearing gloves or visually impaired, leading some researchers to leverage the device motions to design novel gestures for mobile phone interaction. Based on a user’s spatial awareness, Virtual Shelves supports triggering shortcuts mapped to a hemisphere space in front of the body~\cite{Li2009}; however, external sensors for tracking the phone are required. Accelerometer and gyroscope are widely used internal sensors for detecting motion gestures, for instance: auto screen rotation based on a device’s orientation~\cite{Hinckley2000}, rolling a phone on different axes to scroll or command~\cite{Hinckley2012}, tapping the back of a device or jerking it~\cite{Roudaut2009}, and whacking a phone with hard contact forces~\cite{Hudson2010}. Also, through coupling the accelerometer data and touch events on an interactive surface, PhoneTouch allows for direct target selection using the phone and a pick\&drop style of data transferring~\cite{Schmidt2010}.

Combining touch and motion further extends our interaction vocabulary with mobile phones. Hinckley \etal presents a summary of such techniques and how the creation of novel gestures is possible with touch and motion sensors~\cite{Hinckley2011}. Device orientation inferred from motion sensors is one important input to be used simultaneously with touch. TiltText employs tilting angles to resolve text input ambiguities of the traditional phone keypad~\cite{Wigdor2003}. Other usages of tilting and touch include measuring wrist deflection angles~\cite{Rahman2009} and scrolling~\cite{Harrison1998}. ScatterDice Mobile supports the exploration of multi-dimensional data on mobile phones by mapping its orientation to different chart viewing perspectives~\cite{Thomason2012}.

In addition to orientation, device movement can provide more input modalities. For example, Boom Chameleon demonstrates a display tracked with 6 degrees-of-freedom in space as a window in a virtual 3D world~\cite{Tsang2003}. TouchProjector~\cite{Boring2010} and Spilling~\cite{Olsen2007} allow the user to manipulate an object by direct touch on the screen augmented by movement of the phone. Another type of motion involves hard force gestures such as shaking, tapping, and striking the device. Examples include using accelerometer to measure typing pressure~\cite{Iwasaki2009} and the “chucking” technique~\cite{Hassan2009}. Hinckley \etal also present a number of techniques in this category, such as “hold and shake” and “hard drag”~\cite{Hinckley2011}.

In contrast to the above research, our goal is to design techniques from a viewer’s perspective based on the relationship between the device and the user’s face (as opposed to the world position).

There has been some attempt at employing face tracking techniques to provide an extra affordance for mobile interfaces. Kumar \etal utilizes a user’s gaze information for natural scrolling, but additional hardware needs to be added~\cite{Kumar2007}. Hansen \etal\cite{Hansen2006} describes the “mixed interaction space” between the user and the device and proposes using the face to perform image navigation similar to Image Zoom Viewer~\cite{Eriksson2007}. Along the same lines, a face tracking technique is applied in panorama viewing~\cite{Joshi2012} and screen rotation with mobile phones~\cite{Cheng2012}. Recently, head gestures emerged in some off-the-shelf systems, such as Sumsung's “smart screen” technologies~\cite{sumsung} and the accessibility switch control on iOS~7~\cite{ios7}.

However, none of the above work has systematically explored how face movements can be combined with existing touch and motion gestures to augment traditional input methods with new interaction techniques, enabling a more ubiquitous and natural usage for next generation mobile platforms.

\section{Detecting Face-Engaged Gestures}
To enable the engagement of face gestures with touch and motion, we developed a centralized gesture recognizer to collect, process, and integrate various sources of input from the native phone sensors including the capacitive touch screen, accelerometer, gyroscope, and front-facing camera.

All techniques were implemented with an iPhone 5. As provided by the iOS SDK, different touch events such as “touch-beginning”, “touch moving”, and “touch ending” were also employed to detect touch gestures or used for gesture delimiters in some techniques. The accelerometer and gyroscope data were sampled at 60Hz in order to compute phone orientation in real time and detect high-frequency motion gestures such as shaking.

Building on top of the face detection APIs in iOS, we implemented a face input processor to recognize large-angle face rotations, where image sequences of the front camera were sent to it for detecting faces with 0, -45, and 45 degrees of rotations. Based on the detection result sequence, the processor also generates different face events, such as ``face entering", ``face moving", and “face exiting". Face detection and processing was the most computational heavy process in the central gesture recognizer with a frame rate around 16 fps with 480$\times$640 pixel camera image input sequence. Based on our experiments, this face tracking method was adequate for our techniques although using external markers could be faster.

In summary, our centralized gesture recognizer coordinates the input data and events from touch, motion, and face channels as described above at a fixed clock rate, including:

\begin{itemize}
\itemsep -1mm

\item each finger touch screen coordinates, $(T_x, T_y)$, from the touch sensor;
\item three-axis acceleration values, $(A_x, A_y, A_z)$, from the accelerometer;
\item angular rotation rates, $(R_x, R_y, R_z)$, from the gyroscope;
\item and the following parameters via our face input processor from the front camera:
	\begin{enumerate}[a)]
	\itemsep -1mm
	\item face position $(F_x, F_y)$,
	\item face scale $F_s$ estimated from eye distance,
	\item and face angle $F_a$ computed from eye positions.
	\end{enumerate}
\end{itemize}

\begin{figure}[!tb]
\centering
\includegraphics[width=\columnwidth]{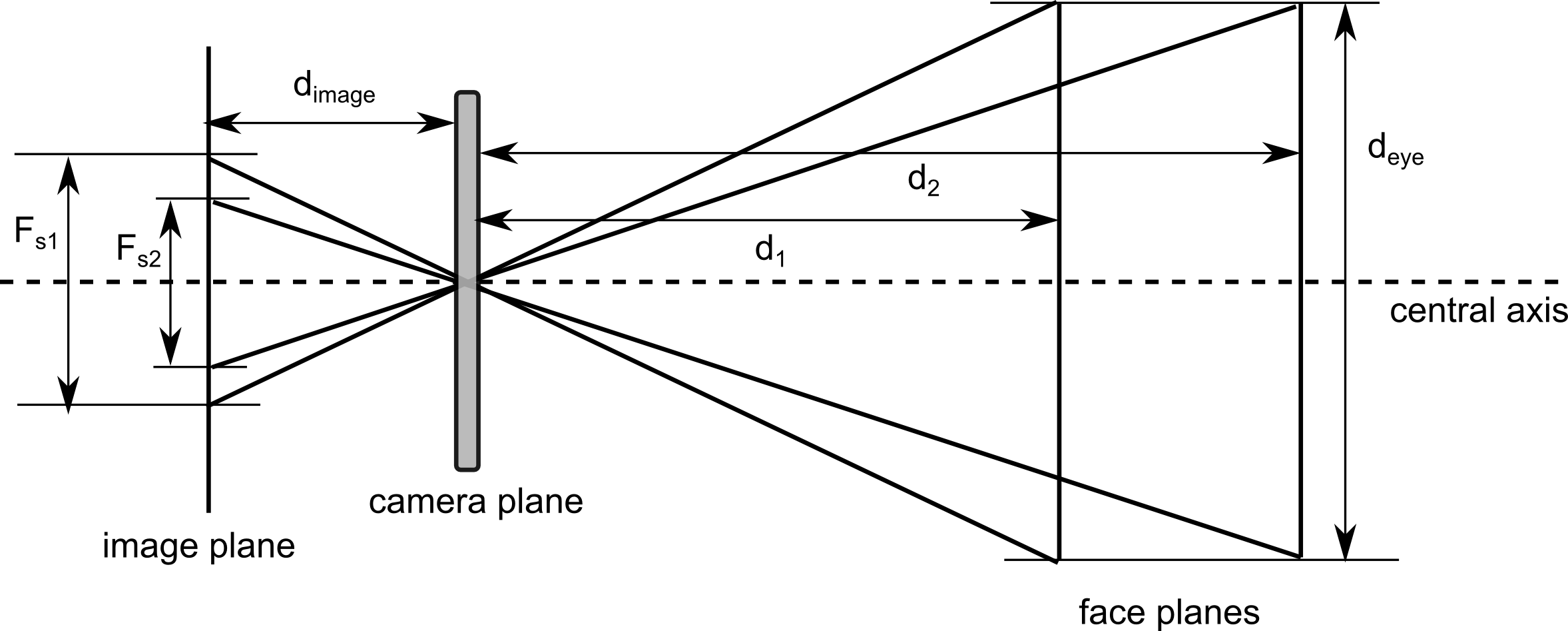}
\vspace{-6mm}
\caption{Representing face-to-phone distance with face scale, where for an individual, $F_{s1}d_1=F_{s2}d_2=d_{eye}d_{image}=constant$.} \label{fig:facescale}

\end{figure}

\begin{figure}[!tb]
\centering
\includegraphics[width=\columnwidth]{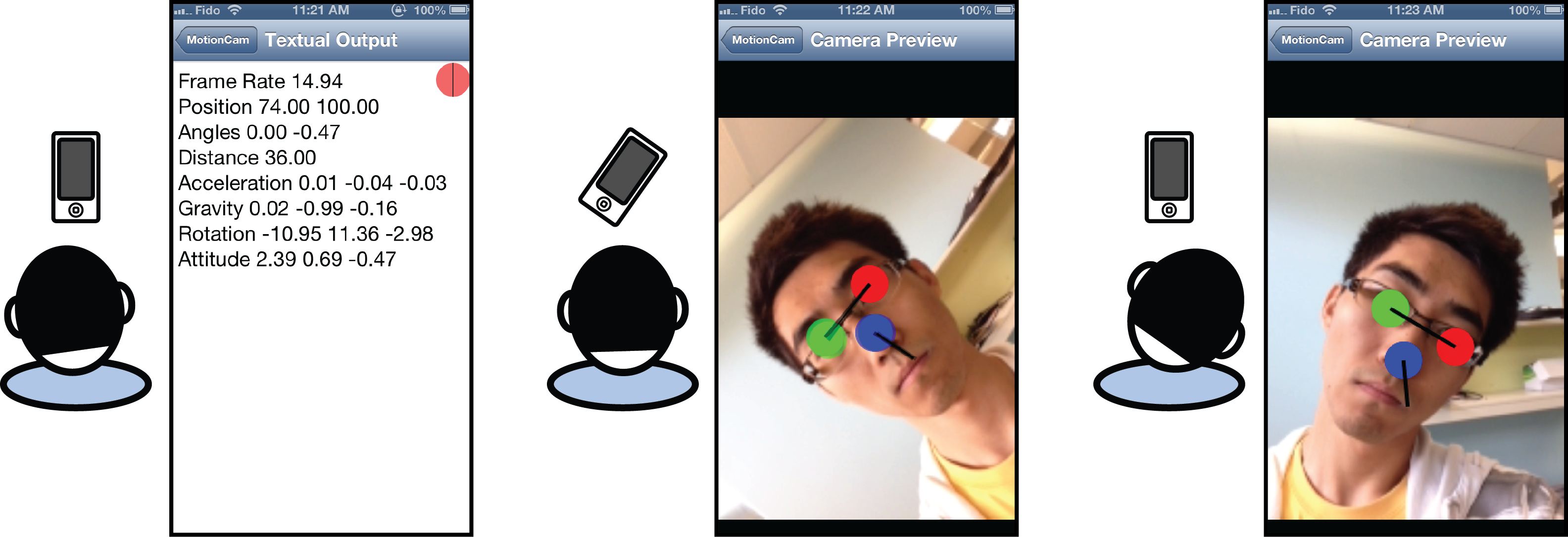}
\vspace{-6mm}
\caption{Sample of debugging outputs.} \label{fig:debug}

\end{figure}

The face scale was used for implying the distance between the head and the phone screen. As per the geometry concept shown in Figure~\ref{fig:facescale}, at any time, the product of face scale Fs and face-to-screen distance $d$ equals to a constant which is determined by one’s eye distance $d_{eye}$ and the camera’s focal length $d_{image}$. Screenshots of the debugging output of the recognizer are shown in Figure~\ref{fig:debug}.

The above face events provide the flexibility for others to build more complicated recognizers, \eg, integrating facial expressions. However, higher level APIs need to be developed to facilitate programmers by encapsulating the above various complicated parameters, in a similar way as gesture recognizers in iOS.

\section{Face-Engaged Interaction Techniques}
We now turn our attention to various techniques enabled by our implementation. We divide the discussion into the following three categories by sensing modality.

\subsection{Techniques: Face-Engaged Touch}
Our first category of techniques combines the touch sensor and the face movement to enhance the traditional touch input for many daily usage scenarios.

\subsubsection{Multi-Scale Scrolling}
Scrolling is necessitated by displaying large content on small screens, and is thus a common interaction performed by mobile phone users when browsing documents, searching in a list, or viewing videos. One important issue in scrolling interfaces is the rate control, i.e., the mapping between virtual content scrolling distance and finger moving distance. The user demands different scrolling speeds depended on the size of content and browsing tasks. For example, the iOS video player uses the perpendicular distance between the touch position and scrollbar to control the scrolling speed for multi-scale navigation (similar to~\cite{Appert2006}). However, it may have hand occlusion problems with touch screens.

We propose a more natural technique that uses the face-to-screen distance to govern the rate of scrolling---the closer the distance, the slower the scrolling speed—using the metaphor that people move text closer to their face if they want to read more carefully. With this technique, users may conveniently adjust the scrolling of content, such as videos and documents, in a spontaneous manner by just moving the device or head. In addition, the representation of content can be modified according to the face-to-screen distance, for example, displaying stock price charts in different scales (by day, week, etc.) while providing different scrolling rates.

\textit{Implementation Details.} Two different multi-scale scrolling mechanisms, absolute and relative scaling, were implemented for this technique. Absolute scaling directly maps the scrolling speed with the quantity of face scale $F_s$, which is another way of inferring the face-to-screen distance without extra sensors. In this method, each face scale value has a fixed scrolling rate. Relative scaling adjusts current scrolling speed based on the relative changes of face scale while the user is actively scrolling. If the device or face is moved during the interaction, the system modifies the scrolling speed based on the direction and distance traveled; and if the user does not perform any interaction, the scrolling speed remains the same even when the face-to-screen distance changes. Through pilot studies, we found that relative scaling was preferred, because users may have various face sizes and different habits of phone holding distances. To improve its stability in situations where the user might not be able to hold the device still, we discretized the possible range of Fs into 6 levels, in which the scrolling speed remained the same. We defined the active scrolling status as: 1) the finger is on the screen, and 2) the time interval between two scrolling actions does not exceed 0.5 second.

\subsubsection{Coarse-to-Fine Text Edit}
Certain text editing tasks can be difficult to perform on smartphone touch screens due to the limited screen space and imprecision of finger input. For example, in cursor positioning, most commercial devices apply the “finger hold” gesture to trigger a virtual magnification lens with fixed offset to the touch point, allowing the user to see beneath their finger (similar to~\cite{Sears1991}). While this is functionally complete, it reduces context of the text surrounding the cursor. Also, it can be difficult for making cursor adjustments near the edges of the screen, because the finger likely slips off the screen thus cannot be sensed.

\begin{figure}[!tb]
\centering
\includegraphics[width=\columnwidth]{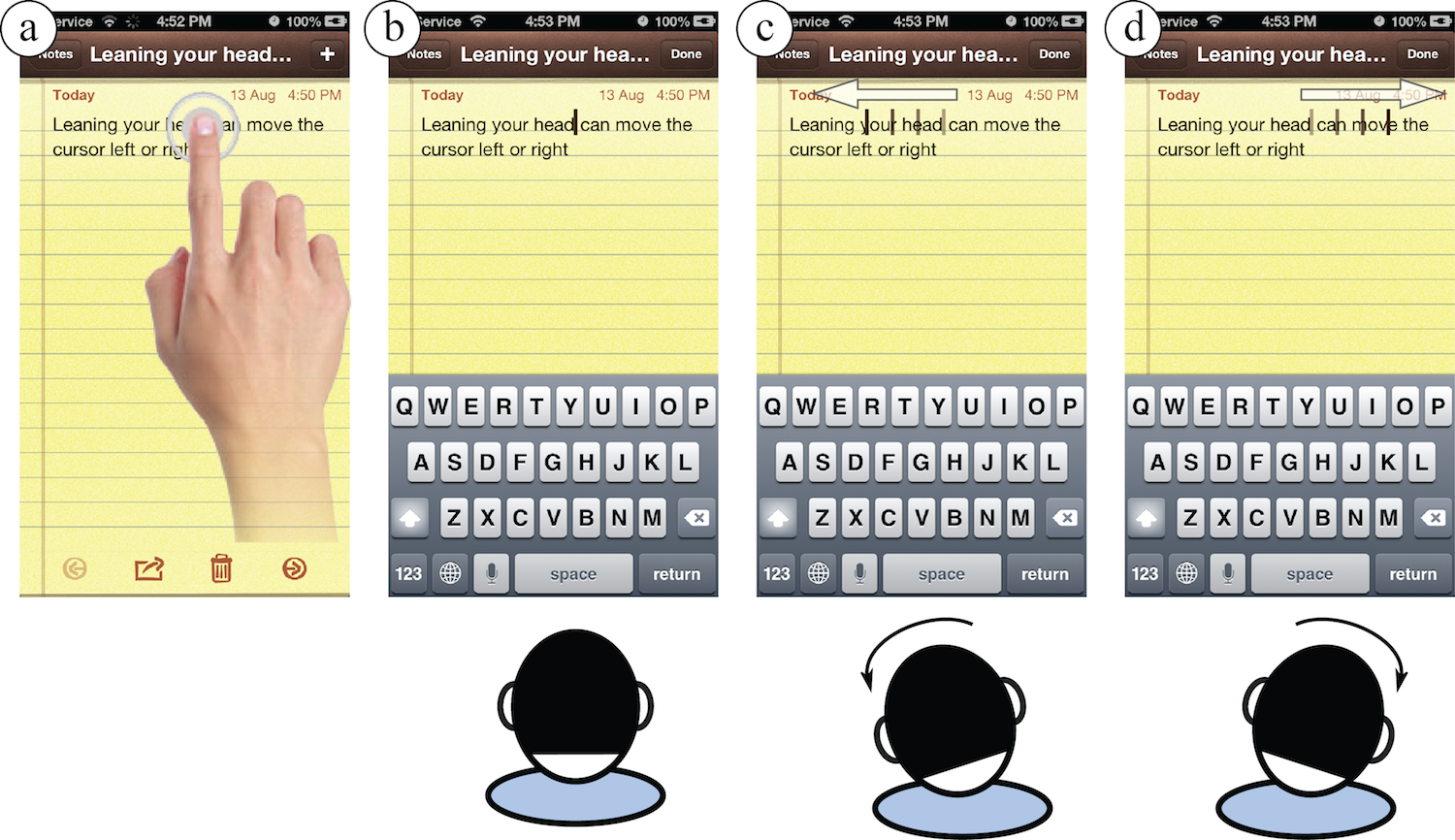}
\vspace{-6mm}
\caption{Coarse-to-fine text edit: (a)(b) first touching the screen to set a rough cursor position, (c)(d) then using head gestures to move cursor in a finer level.} \label{fig:textedit}

\end{figure}

By leveraging the relative orientation between the face and phone, we propose a technique augmenting the classic method of cursor positioning to overcome the above problems (Figure~\ref{fig:textedit}). To place the cursor on the desired location, a user can first tap on the screen normally to give the cursor an approximated position (Figure~\ref{fig:textedit}-ab), and then lean her head left or right to further move the cursor character by character (Figure~\ref{fig:textedit}-cd). Since the second step of fine-level position adjustment does not need touches, it removes the need for the magnifying lens and hence possible occlusions and preserves the context while moving the cursor.

This technique can be embedded into any text editing applications, providing an alternative cursor-manipulation method with minimal hand occlusion. Also note that it does not conflict with the magnifying lens, which can be added onto traditional interfaces to move cursor at different granularities. Further, this technique could also be applied in text selection, where the starting and ending cursors can be manipulated in a similar way.

\textit{Implementation Details.} For activating the face cursor movement, we used a threshold of a 15-degree angle between perpendicular axes of the screen and face. This angle can be computed from the face angle $F_a$ in camera image space and the device screen orientation inferred from the system. During the movement, the cursor shifts at a constant speed (200ms per character, tuned through pilot studies) in the direction controlled by the head, and it stopped moving when the face was in the [-15, 15] degree range. One limitation of this technique may be that a user cannot move the head freely to perform such interaction when side-lying on a bed.

\subsection{Techniques: Face-Engaged Motion}
In addition to face-engaged touch inputs, leveraging sensing of the relative position between the face in the camera image and the device orientation to the ground allows for further interactions.

\subsubsection{3D Map Viewer}
\begin{figure}[!tb]
\centering
\includegraphics[width=\columnwidth]{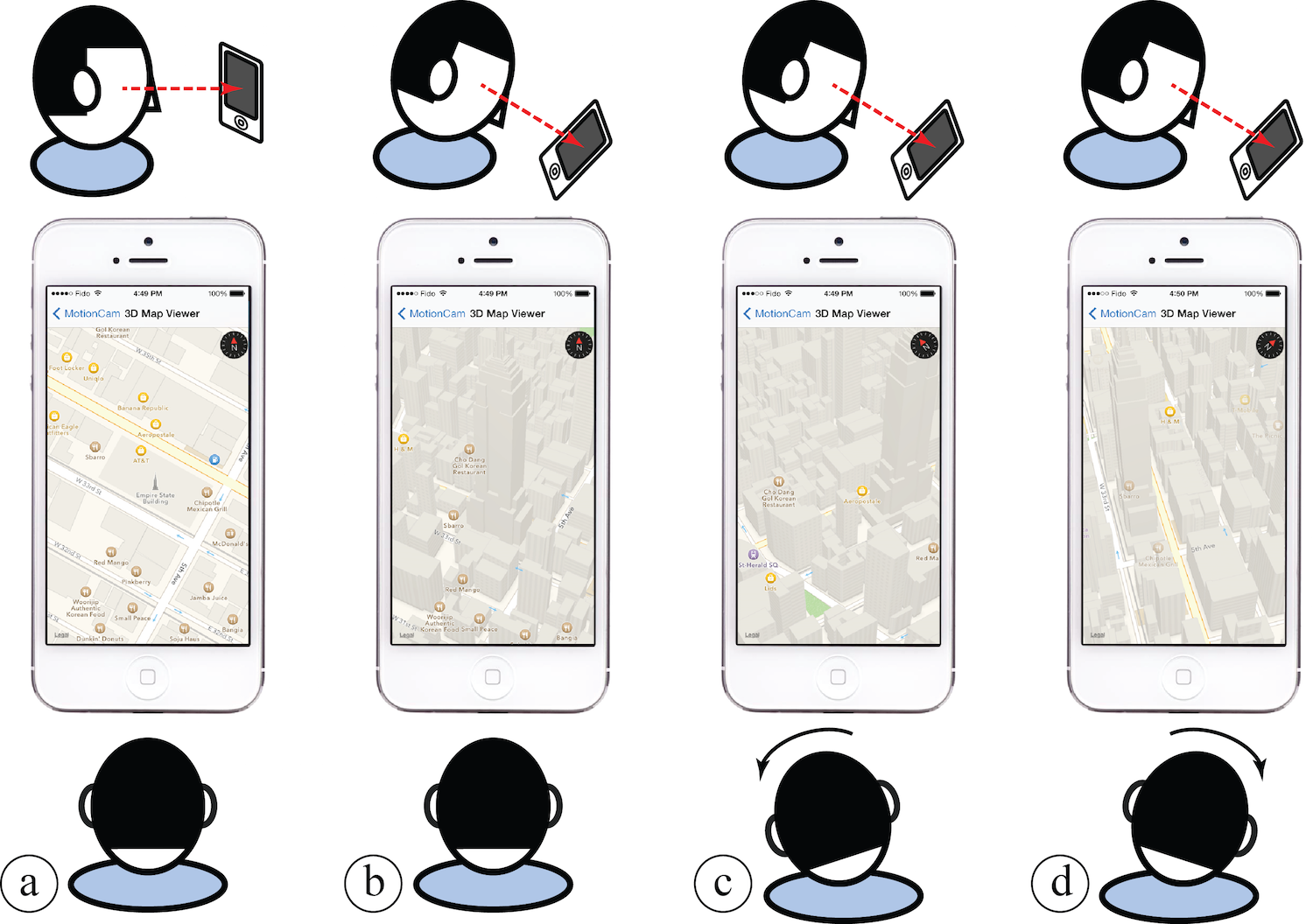}
\vspace{-6mm}
\caption{3D map viewer: (a) normal 2D viewing, (b) rotating the phone to enter the 3D view mode, and (c)(d) moving the head to glimpse left or right side of the 3D buildings.} \label{fig:mapviewer}

\end{figure}

Nowadays mobile geo-map applications usually support not only a traditional 2D map layout but also an immersive 3D view of streets. To have a better navigation experience, people often need to switch between these two views. Current off-the-shelf implementations require users to either explicitly click a “mode” button or drag the map vertically using two fingers. The former is less intuitive and the latter needs a coordination of two-point touch, which is not always possible as discussed before. Thus, we leverage the phone tilting angle to enable a more effective way of transforming between 2D and 3D modes, inspired by our common habits of view perspectives for 3D models. Specifically, a user just naturally rotates the phone to a roughly 45 degree angle relative to the ground from a vertical or horizontal holding of the phone, in order to change from 2D to 3D, or vice versa. Figure~\ref{fig:mapviewer}-ab shows one possible interaction.

In addition, when a user is exploring the 3D map view, it can often be useful to quickly peek the right or left side of the current viewpoint to gain more context of the location. This, however, has not been addressed in any of the current mobile map applications. With the face gestures, a user can rotate the head left or right to control the direction and angle of the glimpse, and if the user rolls the head back to its original position, the view angle goes back to straight ahead (Figure~\ref{fig:mapviewer}-cd), which offers a quick and easy Glimpse-like~\cite{Forlines2005} interaction for 3D map exploration.
When using the above two techniques together, we believe that such face-engaged motion interaction can enhance the user experience of map navigation on mobile phones, which can also be naturally built into the traditional map viewing interactions. Examples of use include virtual sightseeing, first-person perspective gaming, or GPS apps for quick previews of the streets ahead at intersections.

\textit{Implementation Details.} Through iterative design, we have tuned the various parameters to optimize this technique. The initial motion for changing to the 3D view mode is initiated when the titling angle of the device in a range of 45$\pm$10 degrees. To detect head leaning, we employed both face horizontal position $F_x$ and face angle $F_a$ provided by our system. For example, a right leaning gesture was initiated when $F_x$ fell in the right half of the screen while $F_a$ was towards the clockwise direction. This dual-variable approach was used for decreasing the chance for false positives. We further set two minimum thresholds for $F_x$ (80 pixels relative to the camera image center) and $F_a$ (10 degrees relative to the vertical direction) to increase stability. Once the 3D map view was activated by tilting the phone, we used $F_a$ to govern the view peeking angle in the corresponding direction, where three levels were provided at 10, 20 and 30 degrees, mapped to glimpsing angles with 45, 90 and 135 degrees. Similar to the multi-scale scrolling technique, this discretization of the continuous parameter control was intended to incorporate the noisy input.

\subsubsection{Touch-Free Menu}
Using face \& motion input modalities opens the possibility of touch-free interaction. As Zarek \etal\cite{Zarek2012} describe, a number of scenarios of mobile phone use do not allow for capacitive touch input. Such touch-free interaction can be very useful in those scenarios to act as a quick easy manner of achieving necessary tasks. Our goal here is accessory-free input as opposed to leveraging external sensors for tracking, such as~\cite{Li2009}. Though several projects have examined motion sensing techniques using accelerometer, gyroscope, or camera~\cite{Harrison1998,Hassan2009,Hinckley2011,Hudson2010}, few have explored applying face tracking to enable touch-free interaction.

\begin{figure}[!tb]
\centering
\includegraphics[width=\columnwidth]{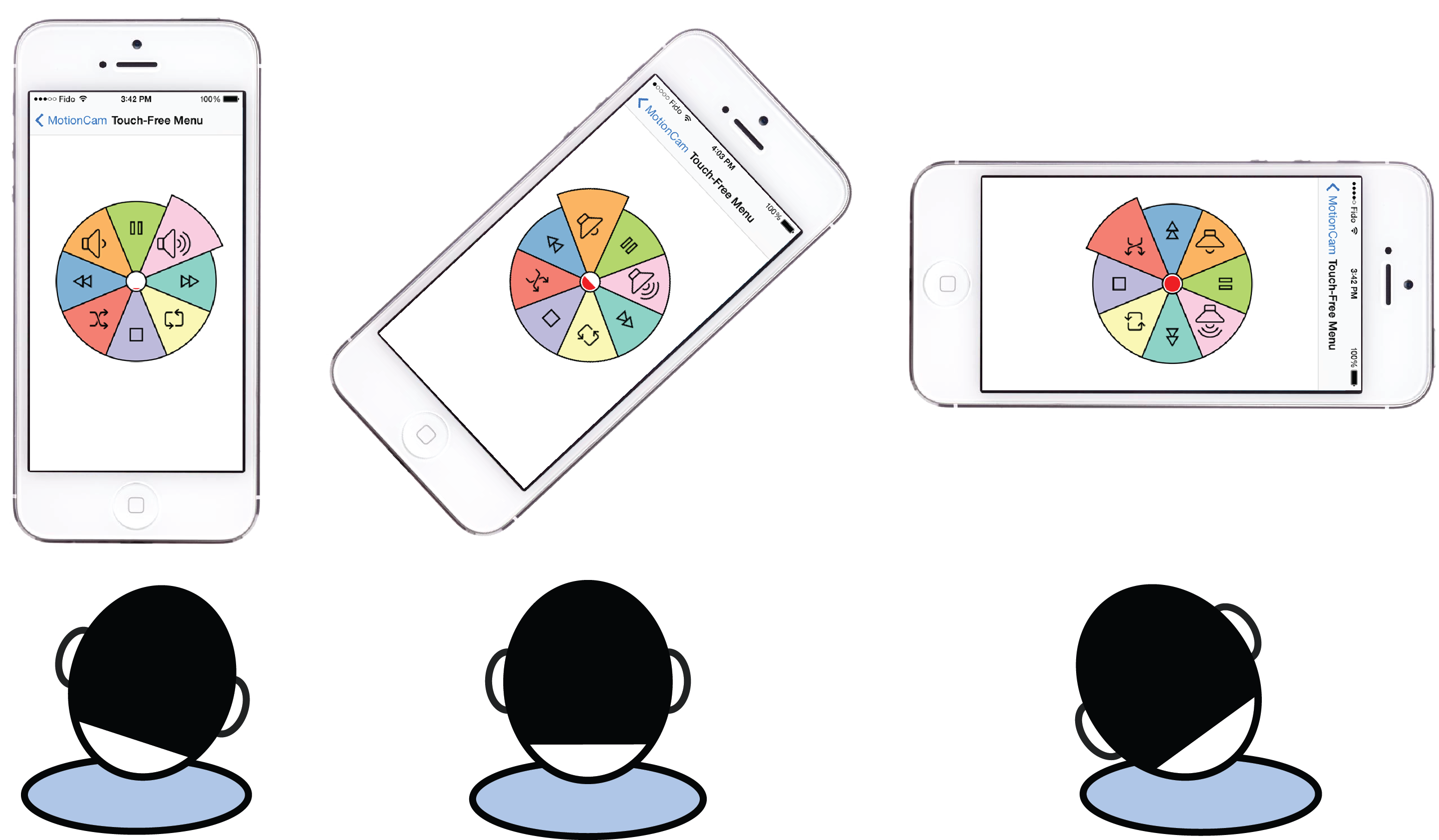}
\vspace{-6mm}
\caption{Touch-free menu: using the relative orientation between the device and face angle to select menu items.} \label{fig:piemenu}

\end{figure}

We propose a pie-menu selection technique by employing the relative angles between the face and device. A user can freely rotate the device or her face to navigate through menu items, where the currently selected item is always aligned with the vertical axis of the face (Figure~\ref{fig:piemenu}). To confirm a selection, a timeout can be applied. Triggering of the pie menu can be contextual (e.g., presented when a phone call is incoming to ``answer on speaker" or ``ignore"), or user-initiated via actuated buttons (e.g., double-clicking the “home” key on an iPhone).

One useful scenario is to augment the music play control on the locked screen with touch-free manipulation of settings during cold weather (e.g., playback, volume, and sound EQ mixing as icons shown in Figure 6). This face-engaged pie menu selection is more flexible than motion-based techniques (e.g., \cite{Li2009}), because it can be used even when a user is lying down and it always keeps the eyes on the screen for displaying additional information.

\textit{Implementation Details.} To demonstrate the basic functions, we implemented a pie menu with 8 menu items, and a selection timeout 3 seconds. Each of these was set through iterative design. We found that 8 items (and thus 45 degrees) was the smallest discretization of the relative orientation of head and phone that could be easily controlled via head motions alone (though turning the phone with the hands did allow for a finer grain). To compute the relative angle of the face to the phone, we used face angle Fa in the image space and the phone orientation inferred from the accelerometer data $(A_x, A_y, A_z)$. We also found that 2 seconds was the fastest timeout that would ensure a very low false activation rate, thus reducing user frustration for accidental selections. However, we can easily employ the face-to-device distance to achieve the selection in this technique in a faster way.

\subsection{Techniques: Face-Engaged Touch and Motion}
Our final combination uses all three input modalities: the movement of the face, touch input, and device motion.

\subsubsection{Expressive Flicking}
Flicking is a widely used interaction on touch screens to enable the use of pseudo-momentum to reduce physical work needed to scroll. However, users sometimes demand richer scrolling interactions for navigating the content with semantics. For example, when reading books, people usually want to jump between chapters or flip pages back and forth, which must be done in multiple steps with normal flicking or tapping gestures.

\begin{figure}[!tb]
\centering
\includegraphics[width=\columnwidth]{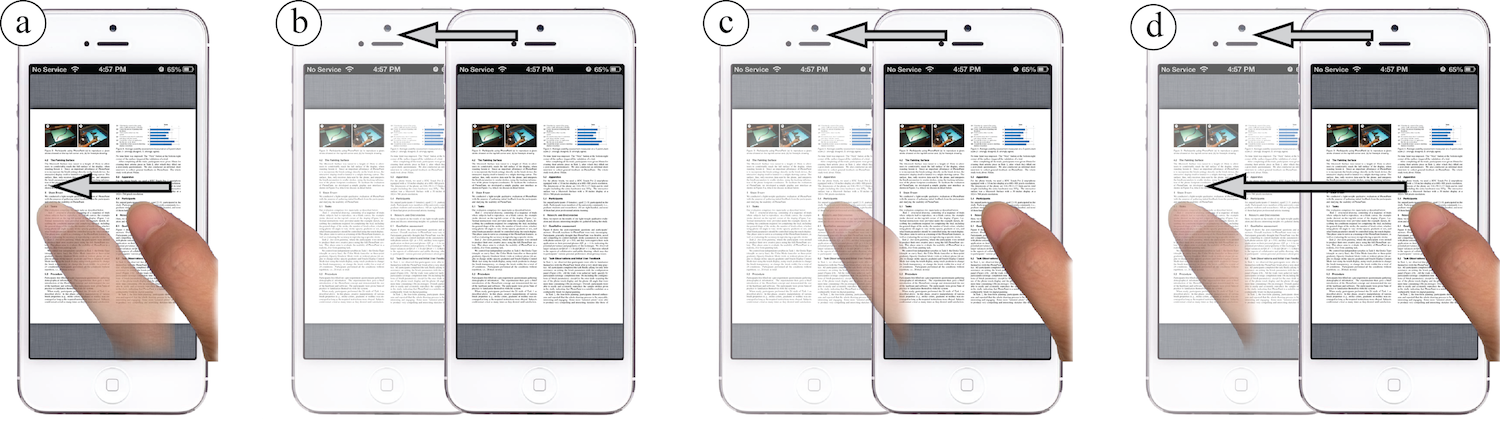}
\vspace{-6mm}
\caption{Expressive flicking: (a) normal flick, (b) phone swipe, (c) hold-and-swipe, and (d) flick-and-swipe.} \label{fig:flicking}

\end{figure}

Through the combination of motion sensing and touch, we introduce expressive flicking, which contains a series of new flicking gestures: phone swipe, hold-and-swipe, and flick-and-swipe (Figure~\ref{fig:flicking}). In phone swipe, the user needs to quickly move the phone left or right starting from the position in front of her face; in hold-and-swipe, the user performs similar phone swipe gestures but with fingers on the touch screen; and in flick-and-swipe, the user needs to initiate a flicking gesture while swiping the phone in the same direction.

Together with the normal flicking, these gestures are ordered in the increasing intensity of actions, which can express different interaction semantics, e.g., scrolling with various levels of distances such as a page, multiple pages, sections, and chapters, for document reading. Compared to the traditional flicking, this technique can improve the efficiency for multi-step tasks with one operation and embed more expressive physical metaphors to interactions.
Unlike the previous techniques designed with only touch and motion, we employed the face detection as a mode indicator for whether to activate the gesture recognizers. More specifically, the flicking gesture is only available when initially performed with the face in front of the screen, to prevent from the user triggering gestures with accidentally touching or shaking the phone.

\textit{Implementation Details.} For implementation, there are some technical considerations. First, one may not be able to hold her finger in the exactly same location when swiping the phone. Thus we added a 15 pixel tolerance for the touch position $(T_x, T_y)$ in hold-and-swipe. The second thing is the synchronization of touch and motion gestures in flick-and-swipe. We looked into the distance traveled by the finger touch from the beginning to the end of a phone swiping, regardless of when the user put down or released her finger, in which the latter could be in the middle of or after the swiping. However, this may result in only utilizing part of the flicking distance for gesture recognition. Lastly, as mentioned above, those gestures (except the normal flicking) must be performed starting with the face being detected, which may not execute the command correctly due to false negatives of face detection, e.g., in a poor lighting condition.

\subsubsection{One-Hand Navigator}
One-hand usage of mobile devices is another common scenario in our daily interactions~\cite{Karlson2008}. Although there have been some attempts to support panning and zooming with one-handed interaction~\cite{Hinckley2011,Joshi2012,Olwal2008,Roudaut2008}, few have addressed rotation, which is commonly used in mapping, image editing and viewing, and graphics design.

\begin{figure}[!tb]
\centering
\includegraphics[width=\columnwidth]{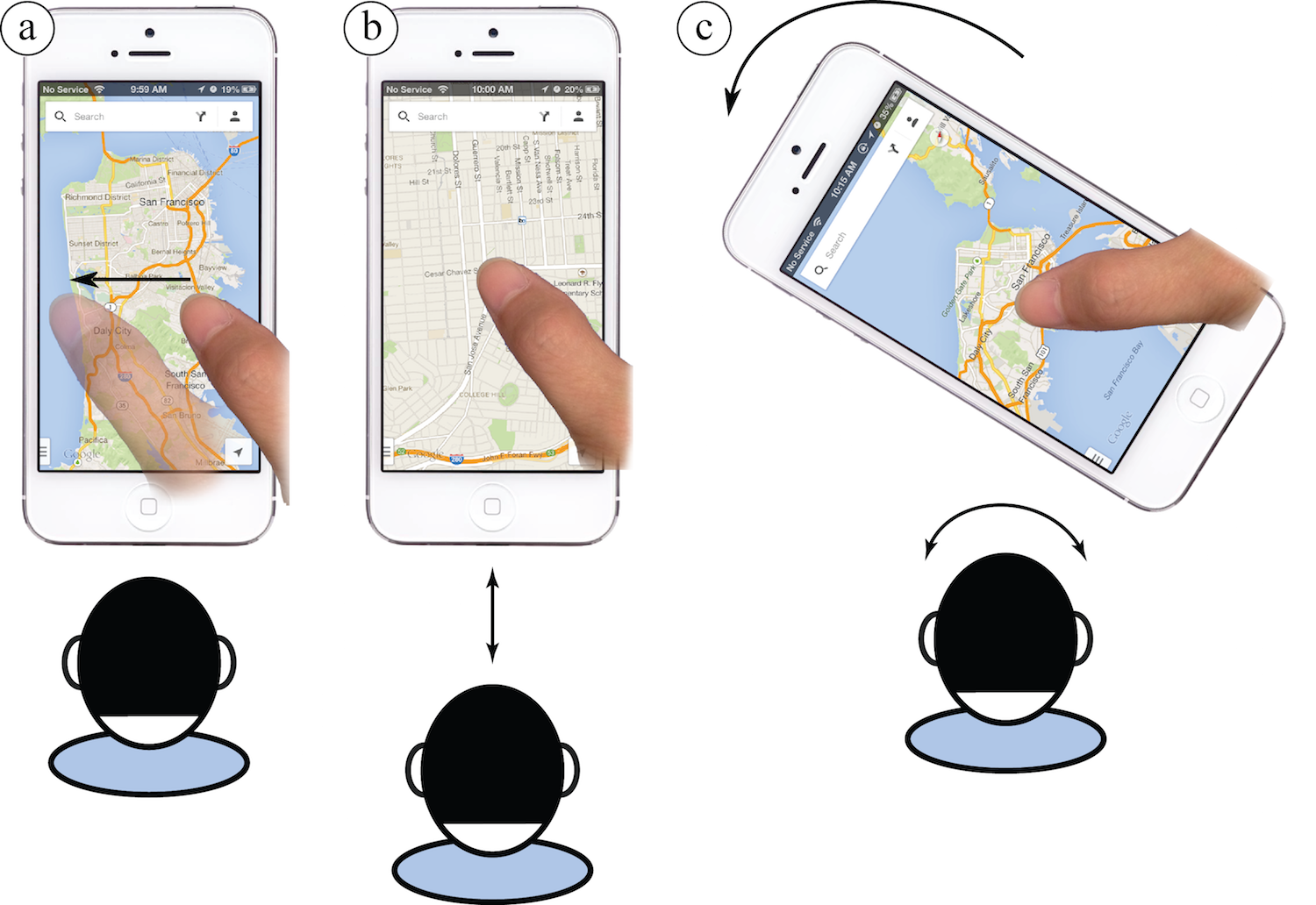}
\vspace{-6mm}
\caption{One-hand navigator: (a) panning with normal finger sliding, (b) zooming by face-to-phone distance, and (c) rotating by face-to-phone orientation.} \label{fig:onehand}

\end{figure}

A demonstration of the technique is shown in Figure~\ref{fig:onehand}. Similar to previous works~\cite{Eriksson2007,Hansen2006}, we utilize the distance between the phone and face to control the zooming scale, and defined the anchor point as the finger touch position on the screen. As discussed in multi-scale scrolling, we apply the relative scaling mechanism for zooming, i.e., the zoom level is adjusted according to difference of the starting and ending face-to-phone distances (Figure~\ref{fig:onehand}-b). Compared to the former techniques (e.g., tilt-to-zoom), such interactions are face-centric which can be executed in any situation even when the user is lying down. To rotate the view, one can put her finger down to set an anchor point and then rotate the device. The original relative orientation between the view and face remains the same while the device is rotating, thus rotating the content onscreen (Figure~\ref{fig:onehand}-c). Similar to zooming, the rotation action stops once the user releases the finger. As for the panning operation, a user can achieve it with the traditional sliding gesture, when zooming or rotation modes are not activated (Figure~\ref{fig:onehand}-a). If the user performs panning during those two modes, only the anchor points are adjusted accordingly.

Compared to prior work, this technique enables one-hand manipulation of maps or images with more completed operations, including pan, zoom, and rotate, all interleaved in a fluid manner. It is useful for cases when a user’s the other hand is occupied that is quite common in our everyday life.

\textit{Implementation Details.} Considering the implementation issues of aforementioned techniques, we applied the same method of discretizing the continuous control in zooming and rotation to reduce the effect of noise. Iterative tuning determined that it was optimal to use 6 levels for zooming in the reasonable range of face scale Fs (that interprets the face-to-phone distance) and 20 levels for rotation in 360 degrees range. Within the same value level where the phone may move slightly, the adjustment of zooming levels or rotation degrees is not performed. However, other advanced rate control techniques (such as in \cite{Joshi2012}) could be employed to provide a smoother user experience.

\section{Initial Evaluation}
Apart from the many small-scale iterative design sessions described throughout the paper, we conducted an initial evaluation to collect user feedback on the six face-engaged mobile interactions. The purpose of this study was to assess the value of these techniques in practice considering both their strengths and weaknesses. We also aimed to observe user interactions and collect their qualitative feedback, since many of the techniques do not have a comparable baseline (\eg, it does not make sense to compare the speed and accuracy between touch selection and touch-free menu). Thus we valued the fluid and convenient user experience in co-existing of the techniques and traditional inputs as well as in special situations that some of the techniques were designed for.

\subsection{Participants}
We recruited 10 participants, including 6 males and 4 females, aged 23-28, all right-handed, from a university network. All participants were daily touch-screen phone users, which would allow us adequately compare our techniques with their everyday experience. Each participant received \$10 as compensation after the study.

\subsection{Apparatus}
Participants were asked to try the proposed techniques using an iPhone 5 weighing 112 grams. The display was a 4.0 inch diagonal with a resolution of 640$\times$1136 pixels. We set its front camera to capture a user’s face continuously at 480$\times$640 pixels resolution during the whole experiment. Custom software was developed for presenting each technique as described in previous sections.

\subsection{Procedure}
For each of the six face-engaged interaction techniques, the experimenter first demonstrated its usage and participants then spent several minutes to get familiar with the gestures and better understand the system. Next, participants were given some instructions to explore and try the features of the technique for about 6 to 10 minutes. After trying each technique, participants filled out a questionnaire of 5 questions using a 1-11 Likert scale (from strongly disagree to strongly agree). The order of techniques presented to participants was randomized. When all the techniques had been explored, we conducted a short informal interview to collect general comments from them, such as the best and worst things of every technique. In the end of the study, participants were asked to rank all six techniques based on their overall preference.

\section{Results}
\subsection{Questionnaires Results}
Overall, the ratings of Likert questions for each technique indicate that participants had positive reactions to the proposed face-engaged interactions with mobile phones (Table~\ref{tab:questionnaire}). Users thought all the techniques were generally easy to learn, where the scores of Q1 were all above 9.0, except for expressive flicking. Some participants commented that they could not distinguish hold-and-swipe and flick-and-swipe in actions: ``\textit{I can easily do another gesture instead}". Some usability issues were observed for techniques requiring coordination of multiple objects simultaneously, such as rotating the phone while moving the head (in touch-free menu) or preforming touch gestures while swiping the phone (in expressive flicking). However, participants felt that it was worth taking some effort to control the techniques through a bit of practice, because they can ``\textit{do a multi-step task with only one move}".

\begin{table}[!tb]
\centering
\small
\setlength{\tabcolsep}{3pt}
\begin{tabular}{lccccc}
\toprule
& \textbf{Q1} & \textbf{Q2} & \textbf{Q3} & \textbf{Q4} & \textbf{Q5} \\ \midrule
\textbf{Multi-scale scrolling} & 10.1 (0.9) & 9.3 (1.7) & 7.7 (1.8) & 8.3 (1.8) & 8.5 (1.8) \\
\textbf{Coarse-to-fine text edit} & 9.7 (1.4) & 8.8 (1.7) & 8.6 (1.7) & 8.2 (2.0) & 7.9 (2.5) \\
\textbf{3D map viewer} & 9.0 (1.8) & 7.4 (1.1) & 7.9 (1.1) & 7.9 (1.1) & 7.4 (1.7) \\
\textbf{Touch-free menu} & 9.9 (1.2) & 9.4 (0.8) & 9.6 (1.1) & 9.4 (0.8) & 9.6 (1.1) \\
\textbf{Expressive flicking} & 7.9 (1.4) & 7.1 (1.3) & 8.6 (0.9) & 8.2 (1.0) & 7.9 (1.4) \\
\textbf{One-hand navigator} & 10.2 (0.8) & 9.9 (1.2) & 10.1 (1.1) & 9.7 (1.2) & 9.7 (1.5) \\
\bottomrule
\end{tabular}
\vspace{-2mm}
\caption{Likert questionnaire results, where in each cell: mean (std). The questions are---Q1: the technique is easy to use, Q2: the technique is easy to learn, Q3: the technique is useful in daily life, Q4: the technique is more efficient than traditional methods, and Q5: I'd like to have the technique on my phone.} \label{tab:questionnaire}

\end{table}

Of all the techniques, one-hand navigator and touch-free menu were rated as the two most useful ones, as some said “this solves many problems in my everyday life”. Users were eager to have those techniques available on their phones. When asked to compare the same task with the standard technique, participants general favored the proposed techniques; especially when only one hand or no hand is available, they agreed that face-engaged interaction techniques were convenient, intuitive, and efficient.

The user preference rankings (Figure 9) indicate the need for supporting effective one-hand and no-touch interactions in our daily life, such as map navigation and menu item selection introduced in this paper. Seven out of eight participants mentioned one-hand navigator as their favorite technique. One user liked multi-scale scrolling the most since ``\textit{it seems very natural to scroll slower when it is far away and it is useful for video browsing}". Expressive flicking was the least favorite technique, because it took so much physical effort and was difficult to control without training. Some participants were also worried that shaking could easily damage their phones.

\begin{figure}[!tb]
\centering
\includegraphics[width=\columnwidth]{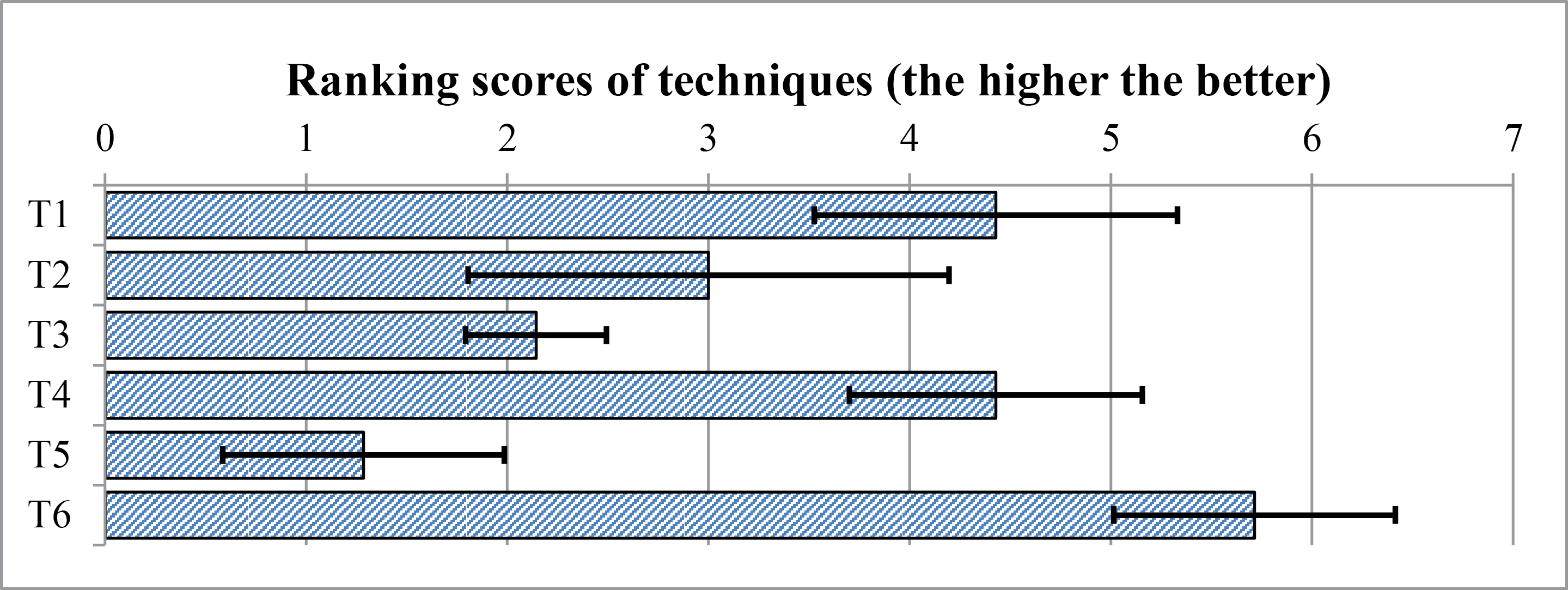}
\vspace{-6mm}
\caption{The average rankings of technique, where T1-6 represents technique in the same order of Table~\ref{tab:questionnaire}.} \label{fig:results}

\end{figure}

\subsection{User Reactions and Observations}
In general, participants enjoyed the process of trying all face-engaged techniques and thought these interactions were useful and natural. They also indicated the experience was smooth and the face tracking was efficient in responding to their face-engaged gestures.

\subsubsection{Learning of Face-Engaged Interaction Techniques}
Most of the techniques seemed to be intuitive, and the real-world metaphors used in many techniques helped users learn the interactions. For example, participants liked the fact that the direction of leaning the head corresponded to the direction of the operation, as moving the cursor in coarse-to-fine text edit. Some mentioned that ``\textit{it makes sense to take more physical effort to achieve more complicated tasks}" in expressive flicking, although it was rated not very easy to learn because users had to manipulate the device simultaneously with touch.

Participants particularly loved one-hand navigator which enables panning, zooming, and rotating of objects all with one hand. They all thought that using the relative face-to-screen distance and orientation to control the zoom levels and rotation angles was very similar to their experiences in the physical world. One suggested that the rotation gesture could be more intuitive if the image orientation remained the same regardless of the head movement. Similar to the zooming mechanism, multi-scale scrolling employs the distance between face and phone to govern the scrolling speed, which was appreciated by most of the participants as well. Only one user mentioned that it was a little strange when directions of the actual scrolling and its speed control were perpendicular.

Another aspect we observed was that users needed to get accustomed to the rate control of parameters in many face-engaged interaction techniques. Due to the noisy inputs of face tracking with the front camera, we set thresholds and discrete levels in the interaction space for head movements to increase the stability despite the sacrifice of continuity. Participants noticed that interface parameter values sometimes appeared jerky and unstable, which required additional time to get familiar with the system settings and some practice to manipulate the interface more precisely. However, integrating advanced face tracking algorithms or the using external sensors may enhance the user experience.

\subsubsection{Everyday Usage vs. Special Situations}
All participants agreed that the proposed techniques were useful under certain situations, as our goals are to augment traditional interactions rather than replace them. For expressive flicking, some users indicated that they would use it for browsing a large number of images or extensive documents, but normally they do not perform those tasks on mobile phones. One participant intended to apply coarse-to-text edit for texting: ``\textit{it releases my hands to do other stuff and brings the cursor back to the desktop [manner]}".

Some participants even indicated that they would like to use multi-scale scrolling, touch-free menu, and one-hand navigator in daily interactions. One commented that ``\textit{[one-hand navigator] is easier to mix zooming and rotation. […] I want to have it for my Google Map}". Another user added a number of scenarios for a touch-free menu, such as when she is cooking and her hands are dirty. Some said that multi-scale scrolling could be very useful on tablets for reading books or skimming online videos.

\subsubsection{Ergonomics, Physical Effort, and Social Awkwardness}
Although face-engaged interaction techniques were generally considered natural and intuitive by participants, the form of griping the device, the motion range of wrist and the movement space of head should be carefully considered into an appropriate gesture design.

Participants thought that it was much easier to rotate the phone than rotate the head. Some suggested that the touch-free pie menu should be designed as 270 degrees instead of a full circle, as ``\textit{it is hard to choose the lower end [of the menu]}". Given much freedom, participants were sometimes confused about whether to rotate their heads or the phone to complete a task. We also observed that some users felt a little uncomfortable holding the head stable while rotating, e.g., for coarse-to-fine text edit. But participants indicated that they could manage it, because they would use such interactions for special situations which are not likely to happen all the time.

Moreover, several users thought that the benefits of doing complicated tasks in one single action might be decreased due to the physical effort needed for the gesture, e.g., flick-and-swipe in expressive flicking. Participants said that in most of the cases, the traditional ways might be preferred though it took longer, despite the fact that they liked the physical metaphors, interaction semantics, and intuitions applied in the techniques.

Another issue revealed during the interviews was that people might feel awkward when performing the face-engaged interaction techniques in front of others, especially those requiring the rotation of the head (e.g., touch-free menu). Also, one mentioned that when it was crowded, there might not be enough space to do actions such as swiping the phone and moving the phone far from the face.

\section{Discussion}
From the study, although these face-engaged techniques were considered to be intuitive and convenient in many usage scenarios, they limit interactions to a restricted space where the face must be captured and tracked. In our system, the user has to be front-facing and hold the phone within a certain distance ranges. However, having a wider view-angle camera may solve these issues. During the experiment, we did not limit users with certain standing postures or facing directions, and the sensing seemed relatively stable throughout the study. However, lighting conditions might be different for outside environments, which could affect the tracking.

Moreover, some face-engaged gestures that require frequent head rotations (e.g., coarse-to-fine text edit) might have the fatigue issue. However, we argue that those techniques are not specially designed for daily usage but for just-in-time and auxiliary use to achieve tasks in special scenarios that normally cannot be done.

Another limitation is that the face tracking input is usually noisy (especially in non-stationary environments) and has huge latency, thus it cannot be used for precise and high-frequency parameter adjustment. Also, false positives or the loss of face tracking may interrupt the interaction, especially in continuous control. Nonetheless, from the experiments, our current implementations together with some technique specific adjustments (\eg, discretization of continuous parameter, and dual variable gesture detection) seem to be adequate for simple everyday tasks. But the development of better face tracking algorithms and faster hardware on the phone would significantly enhance the user experience. It is also interesting to implement temporal filtering techniques such as Kalman filter~\cite{Kalman1960} to stable the input.

From the observations and interviews, we also identified an interesting point about the trade-off of speed and convenience versus effort and learning. While these face-engaged gestures provide a rich interaction vocabulary to do complicated tasks more efficiently, users need practice to best manage the new interactions. Sometimes users have to take more physical effort, such as in expressive flicking, to overcome the inefficiency of common techniques. People may not always prefer face-engaged interactions on a daily basis, but many of our techniques are designed to augment traditional inputs, especially under unusual situations, e.g., when another hand or touch is not available.

\section{Conclusion and Future Directions}
We have explored novel interaction techniques for mobile devices by augmenting the existing touch and motion gesture paradigm with face movements, which facilitates many special usage scenarios in our daily life. Three groups, in total six techniques, were discussed based on their sensing methods, including face-engaged touch, face-engaged motion, and face-engaged touch \& motion. From user reactions in the study, we conclude that these new techniques are intuitive and efficient in many situations compared to the traditional methods, and have indicated a richer interaction vocabulary with extra affordances. Further, by framing the previous work and those techniques in a conceptual structure based on the parameter control of the three input modalities, we believe that these face-engaged interactions can provide inspirations and implications for future research to better exploit face input, the underutilized channel.

In the future, we would like to conduct more experiments and further extend the highly-rated techniques. It is also interesting to continually populate the design space by designing and testing other novel face-engaged interaction techniques. Finally, it is promising to enhance the current gesture recognizer by exploring more advanced face tracking algorithms and the use of other sensors.


\bibliographystyle{abbrv}
\bibliography{motionspace}
\end{document}